\documentclass[10pt,a4paper,twoside,twocolumn,english,british,aps,prd, nofootinbib,preprintnumbers]{revtex4-1}
\usepackage{lmodern}

\usepackage[T1]{fontenc}
\usepackage[utf8]{inputenc}
\usepackage{color}
\usepackage{babel}
\usepackage{amsmath}
\usepackage{amssymb}
\usepackage{graphicx}
\usepackage[unicode=true,pdfusetitle,
 bookmarks=true,bookmarksnumbered=false,bookmarksopen=false,
 breaklinks=true,pdfborder={0 0 0},backref=false,colorlinks=true]
 {hyperref}

\makeatletter

\pdfpageheight\paperheight
\pdfpagewidth\paperwidth

\@ifundefined{textcolor}{}
{%
 \definecolor{BLACK}{gray}{0}
 \definecolor{WHITE}{gray}{1}
 \definecolor{RED}{rgb}{1,0,0}
 \definecolor{GREEN}{rgb}{0,1,0}
 \definecolor{BLUE}{rgb}{0,0,1}
 \definecolor{CYAN}{cmyk}{1,0,0,0}
 \definecolor{MAGENTA}{cmyk}{0,1,0,0}
 \definecolor{YELLOW}{cmyk}{0,0,1,0}
}

\renewcommand\[{\begin{equation}}
\renewcommand\]{\end{equation}} 
\usepackage{hyperref}
\hypersetup{
    colorlinks,%
    citecolor=blue,%
    filecolor=blue,%
    linkcolor=blue,%
    urlcolor=blue
}

\makeatother

\begin{document}

\preprint{CERN-PH-TH/2012-242}

\title{Hidden Negative Energies in Strongly Accelerated Universes}

\author{Ignacy Sawicki}

\affiliation{Institut für Theoretische Physik, Ruprecht-Karls-Universität Heidelberg,\\
Philosophenweg 16, 69120 Heidelberg, Germany}

\email{ignacy.sawicki@uni-heidelberg.de}

\selectlanguage{british}%

\author{Alexander Vikman}

\affiliation{CERN, Theory Division, CH-1211 Geneva 23, Switzerland, }

\affiliation{Department of Physics, Stanford University, Stanford, CA 94305, USA}

\email{alexander.vikman@cern.ch}

\date{\today}
\selectlanguage{british}%
\begin{abstract}
We point out that theories of cosmological acceleration which have
equation of state, $w$, such that $1+w$ is small but positive may
still secretly violate the null energy condition. This violation implies
the existence of observers for whom the background has infinitely
negative energy densities, despite the fact that the perturbations
are free of ghosts and gradient instabilities. 
\end{abstract}
\maketitle

\section{Introduction}

One frequently posits that dark energy (DE) is described by a perfect-fluid energy-momentum
tensor (``EMT''), 
\begin{equation}
T_{\mu\nu}=\left(\varepsilon_{u}+P\right)u_{\mu}u_{\nu}-g_{\mu\nu}P\,,\label{eq:EMTBackground}
\end{equation}
where $u^{\mu}$ is the velocity of an observer while
$\varepsilon_{u}$ is the energy density and $P$ is the pressure
measured by the observer and we use the $(+---)$ convention for the
metric. The equation-of-state parameter is then defined as $w\equiv P/\varepsilon_{u}$. 

In cosmological setups, one
then considers homogeneous and isotropic Friedmann universes. In such a case, even for more general, imperfect-fluid DE models, the configuration of the background EMT takes the form of Eq.~\eqref{eq:EMTBackground} with $u^\mu$ taking on the meaning of the cosmological observer. The Universe accelerates for $w<-1/3$ and does so more strongly the
more negative is $w$. The cosmological constant, with $w=-1$ exactly,
is compatible with all the observations. However, the data show, see
e.g. \cite{Komatsu:2010fb,Suzuki:2011hu,Hinshaw:2012fq}, a statistically insignificant
preference for DE with a slightly more negative equation of state
for at least a part of the history of the universe. Moreover, given
realistic future surveys, even if $w$ is constant, the measurement
error will remain at 2\%, see e.g.~\cite{Amendola:2012ys}. The possibility
of strong acceleration, $w<-1$, sometimes called a \textit{phantom} \cite{Caldwell:1999ew},
is striking, because it implies a that the EMT violates the \textit{null energy
condition} (``NEC''), which states that for \textit{all} null vectors
$n{}^{\mu}$ and $T^{\mu\nu}:$ $T_{\mu\nu}n^{\mu}n^{\nu}\geq0$ and
for the EMT given by (\ref{eq:EMTBackground}) is equivalent to $\varepsilon_{u}+P\geq0.$
This is the weakest of all classical energy conditions and its violation
may have dramatic implications for the future of the universe \cite{Starobinsky:1999yw,Caldwell:2003vq}. 

The question of whether it is at all possible to violate the NEC without
any perturbative instabilities has only been answered recently. Perfect
fluids and canonical scalar fields which violate the NEC \emph{always} suffer
from either ghosts or gradient instabilities or both, see e.g. \cite{Dubovsky:2005xd}. These are short-timescale
instabilities that are most rapid at energies near the cut-off and
therefore immediately disqualify the background solution, see however
\cite{Emparan:2005gg,Garriga:2012pk}. In Refs \cite{Nicolis:2009qm,Creminelli:2010ba,Deffayet:2010qz,Kobayashi:2010cm}
noncanonical scalar-field models were constructed that can serve as
a medium with $w<-1$ and where perturbations have a positive kinetic
term and real sound speed. These scalar fields belong to theories
which contain higher-order derivatives in the action but retain second-order
dynamical equations \cite{Horndeski,Deffayet:2011gz,Kobayashi:2011nu}
extending the so-called \textit{galileons} \cite{Nicolis:2008in,Deffayet:2009wt,Deffayet:2009mn,deRham:2010eu}.
A feature shared by all these models is that, contrary to scalar fields
without higher derivatives in the action, these models do \emph{not
}have EMTs of perfect-fluid form \eqref{eq:EMTBackground} on generic
configurations.

In this paper, we point out two peculiarities of DE configurations
which violate the NEC. First we  provide a general proof that a violation
of the NEC by \emph{any matter} on some \emph{arbitrary} (not necessarily homogeneous or isotropic) configuration implies that the corresponding energy density as a function in phase space at a given point is necessarily unbounded from below. This is true even if there are no ghosts or gradient instabilities around this configuration.  Thus the existence of a single NEC-violating configuration always implies that there are other configurations of this matter on which the Hamiltonian is arbitrarily negative, if, as is usually the case, the Hamiltonian is equal to the energy density. As a consequence, observers boosted with speed higher than $\left(-w\right)^{-1/2}$
with respect to the cosmological background see negative energy
densities of the NEC-violating DE, the more negative the faster they move, without any
limit. Moreover, these observers see the flows of DE energy density
as spacelike. In particular, given the current knowledge of $w$,
this may apply to e.g.\ massive particles in cosmic rays. 

Further we point out that even measuring $w>-1$ does not preclude
the possibility of a violation of the NEC by means of the purely spatial
parts of the EMT: energy flows and anisotropic stresses which can
be relevant in perturbations around strongly accelerated backgrounds
with positive but small $1+w$. As we have mentioned above, the
only known systems violating the NEC without immediate pathologies
in perturbations necessarily possess such imperfect EMTs. Thus again
there would be observers measuring infinitely negative energy densities
of DE. 

All of this applies equally well to NEC-violating models of the early
universe, e.g.\  \cite{Creminelli:2010ba,Kobayashi:2010cm,Qiu:2011cy,Easson:2011zy}.

Whether the unbounded negative energy densities we discuss in this paper definitely imply a problem, we are unsure. However, their existence is definitely puzzling and somewhat uncomfortable. 

\section{$w<-1$ implies unbounded negative energies}

The violation of the NEC for matter with the EMT $T_{\mu\nu}$ is
defined as the existence of at least one future-directed ray of null
vectors $n^{\mu}$ such that $T_{\mu\nu}n^{\mu}n^{\nu}<0$ . In a
frame $u^{\mu}$, the observed energy density $\varepsilon_{u}=T_{\mu\nu}u^{\mu}u^{\nu}$.
Below we show that in that case, there always exists a frame $V^{\mu}$
where $\varepsilon_{V}=T_{\mu\nu}V^{\mu}V^{\nu}<\varepsilon_{u}$.
In particular, for any positive $\varepsilon_{u}$, there are frames
where $\varepsilon_{V}<0$. Let us prove this statement. 

Let us chose the null vector from the NEC-violating ray such that
$n^{\mu}u_{\mu}=1$. We parameterise the new frame as 
\[
V^{\mu}=\alpha\, u^{\mu}+\eta\, n^{\mu}\,.
\]
This vector is timelike provided 
\begin{equation}
\alpha^{2}+2\alpha\eta=1\,,\label{eq:Normalization}
\end{equation}
and the new frame is future-directed when $V^{\mu}u_{\mu}>0$. The
frame $V^{\mu}$ moves with respect to the frame $u^{\mu}$ with the
speed 
\begin{equation}
v=\frac{\eta}{\alpha+\eta}=\frac{1-\alpha^{2}}{1+\alpha^{2}}\,.\label{eq:BoostVel}
\end{equation}
Substituting $\eta$ from (\ref{eq:Normalization}) we obtain
\begin{align}
\varepsilon_{V}\left(\alpha^{2}\right)=\alpha^{2}\varepsilon_{u}+\left(1-\alpha^{2}\right)\, T_{\mu\nu}n^{\mu}u^{\nu}+\label{eq:AlphaEnergyDensity}\\
+\frac{1}{4\alpha^{2}}\left(1-\alpha^{2}\right)^{2}T_{\mu\nu}n^{\mu}n^{\nu}\,,\nonumber 
\end{align}
so that 
\begin{equation}
\frac{\partial^{2}\varepsilon_{V}}{\partial\alpha^{2}\partial\alpha^{2}}=\frac{1}{2\alpha^{6}} \ T_{\mu\nu}n^{\mu}n^{\nu}\ .
\end{equation}  
Thus the energy density as a function of $\alpha^2$ can have a minimum only if the NEC holds.
Indeed,  in frames very close to the NEC-violating vector $n^{\mu}$,
i.e.\ for small $\alpha$, the energy density $\varepsilon_{V}\left(\alpha^{2}\right)$
behaves as 
\begin{equation}
\varepsilon_{V}\left(\alpha^{2}\right)\simeq\frac{1}{4\alpha^{2}}\, T_{\mu\nu}n^{\mu}n^{\nu}<0\,,\label{eq:eV_negative}
\end{equation}
and can be made arbitrarily negative. 
\\

We can now reinterpret the effect of this passive Lorentz transformation, as an active one: instead of observing the same configuration from different reference frames we can change the configuration itself. By virtue of Lorentz symmetry, these boosted configurations are solutions as well and therefore belong to the phase space. Thus if a boosted observer can observe arbitrarily negative energy density, this implies that the phase space contains configurations with arbitrarily negative energy, i.e.\ that the Hamiltonian is unbounded from below. 
There are two types of systems for which this conclusion may not be applicable: (i) systems without fundamental Lorentz invariance, (ii) systems where the Hamiltonian does not coincide with the energy density given by the EMT.   
\\

Let us now assume that the cosmological background of DE has a positive
energy density, $\varepsilon_{u}$, in (\ref{eq:EMTBackground}) and
violates the NEC only slightly: namely $w<-1$ but so close to $-1$
that it may remain unobservable for future probes. Further the perturbations
are not ghosty and have positive sound speed squared, $c_{\text{s}}^{2}>0$,
i.e.\ this DE seemingly does not suffer from any perturbative instabilities.
Let us now boost away from the cosmological frame with speed $v$
in any direction. Then Eq.~(\ref{eq:AlphaEnergyDensity}) implies
that the energy density of DE in the new frame, $\varepsilon_{V}$,
becomes negative when 
\[
v>v_{\text{crit}}=\frac{1}{\sqrt{-w}}\,.
\]
Inside structures the total EMT is dominated by dark matter (``DM'')
and therefore the \textit{total} energy density measured by \textit{all}
observers will still be positive. However, for a sufficiently smooth
DE there are large voids where DE dominates the total EMT over DM. 

Now let us recall that the energy of the protons at the LHC is $4$
TeV so that $1-v_{\text{p}}\simeq10^{-7}$, whereas cosmic rays have
been observed up to energies of order $10^{19}$ eV \cite{Settimo:2012zz},
giving $1-v_{\text{cray}}\simeq10^{-20}$, assuming that cosmic rays
consist of protons. 

This implies that in order that the energy density of the cosmological
background of DE be positive in the rest frame of observed particles,
\[
1+w\gtrsim-10^{-20}\,.
\]
This is clearly a strong requirement far beyond the precision of any
future experiments. Even stronger conditions can be obtained from
ultra-high energy neutrinos e.g. by the IceCube detector \cite{Abbasi:2010ie}
where neutrino up to 400 TeV were measured which gives roughly $v_{\nu}\simeq1-10^{-28}$.

In addition, the density of the fluid's 4-momentum as measured by
an observer moving with $V^{\mu}$ is $p_{\mu}=T_{\mu\nu}V^{\nu}$.
This vector becomes \textit{spacelike} when $v>1/\left|w\right|$,
thus at even smaller observer speeds than those corresponding to negative
$\varepsilon_{V}$. Under normal circumstances, the spatial part of
$p_{\mu}$ points against the direction of the boost potentially providing
a force to resist further acceleration. When the NEC is violated,
the opposite is true, seemingly aiding the motion. 

At this point we should ask whether this peculiar observation necessarily
signifies a problem. This should be a question of the structure of
interactions. If there are no direct interactions between usual particles
and the DE, except through gravity, then it is unlikely that any process
could occur sufficiently quickly to have an effect. However, one should
keep in mind that, if it is possible to extract this negative energy
density, the particle could be boosted even further and see an even
more negative energy density of the background. Also this could lead
to violations of the second law of thermodynamics and run-away instabilities. 

The situation is more interesting for small fluctuations of DE itself
-- ``phonons''. They definitely interact with the background. If
their maximal speed of propagation -- i.e.\ the sound speed $c_{\text{s}}$
-- lies in the interval $v_{\text{crit}}<c{}_{\text{s}}<1$ then these
fluctuations in their own rest frame interact with the negative energy
density of their own background. Since we have required that the perturbations
be healthy as it is in e.g.\  \cite{Deffayet:2010qz}, the phonons'
energy is positive and they behave like normal particles. 

If the presence of this negative energy density is a problem it would
result in the pumping of energy into phonons making the background
less and less homogeneous. 

NEC-violating theories can also have interesting applications in models
of the early universe. In particular, G-\textit{inflation} \cite{Kobayashi:2010cm}
can create a blue spectrum of gravitational waves, while models \cite{Creminelli:2010ba,Qiu:2011cy,Easson:2011zy}
provide insight on the initial state of the universe. It is interesting
to note that in the particular case studied in detail in \cite{Kobayashi:2010cm}
the sound speed is $c_{\text{s}}\lesssim0.18$, while $w+1$ can be
made arbitrary small therefore avoiding these issues.

In the case of \textit{Galilean Genesis} \cite{Creminelli:2010ba},
at the beginning of the universe, $\tau\rightarrow-\infty$, one has
$c_{\text{s}}\simeq1-\frac{8}{3}\tau^{-2}$ and $w\simeq-1-4\tau^{2}$
corresponding to $v_{\text{crit}}\simeq\frac{1}{2}\tau^{-1}\ll c_{\text{s}}$
so that phonons in their rest frame interact with a background of
a very negative energy density. 

Here one should keep in mind that to bounce the universe one has to
require that the energy density of the NEC-violating medium in the
cosmological reference frame at least vanishes or is even negative
to compensate for the presence of the normal matter, see \cite{Easson:2011zy}.
However, this is only necessary around the bounce / turnaround point.
For vanishing energy density of the medium the above argument would
imply that $v_{\text{crit}}=0$ at that moment. 

To finish this section on a more hopeful note we would like to remark
that if our observation can signify a stability problem for the above
models, it could also be considered as a feature allowing for a defragmentation
of the inflaton background, i.e.\ reheating to end inflation.

\section{Violating the NEC with $w>-1$}

All models known so far which violate the NEC and have healthy perturbations
possess an EMT which is \emph{not} of the perfect-fluid form (\ref{eq:EMTBackground})
on \textit{realistic} cosmological configurations which are locally
slightly inhomogeneous and anisotropic. The simplest such example
is \textit{imperfect} DE from a scalar field $\phi$ with \textit{Kinetic
Gravity Braiding} \cite{Deffayet:2010qz} which for all timelike $\partial_{\mu}\phi\propto u_{\mu}$
has EMT 

\begin{equation}
T_{\mu\nu}=\left(\varepsilon_{u}+P\right)u_{\mu}u_{\nu}-g_{\mu\nu}P+u_{\mu}q_{\nu}+u_{\nu}q_{\mu}\,,\label{e:EMT}
\end{equation}
corresponding to an \textit{imperfect} fluid \cite{Pujolas:2011he}.
Usually the purely spatial $q_{\mu}$, orthogonal to the frame $q^{\mu}u_{\mu}=0$,
is attributed to heat flow and entropy production. However, in case
of \cite{Deffayet:2010qz,Pujolas:2011he} it is just the flow of charges,
resulting from the fact that $u^{\mu}$ does not correspond to the
Eckart frame. The four-velocity $u^\mu$ is uniquely chosen by requiring there be no anisotropic stress. 

On large scales, the spatial vector $q_{\mu}$ averages
to zero. However, locally it contributes to flows of DE perturbations.
On average, $u^{\mu}$ coincides with the cosmological frame. Note
that more general scalar-field models violating the NEC would necessarily
possess anisotropic stress, see e.g.\ \cite{Barreira:2012kk}. Motivated
by the above example, below we consider DE systems with EMT given
by \eqref{e:EMT} without making any reference to the underlying theory. 

Unsurprisingly the restrictions put on four-scalars $P$, $\varepsilon_{u}$ and the four-vector $q^{\mu}$ by the various energy conditions differ from those for the perfect fluid. In particular $T_{\mu\nu}n^{\mu}n^{\nu}$ is minimal
on the ray with $n^{\mu}=u^{\mu}+q^{\mu}/q$ where $q=\sqrt{-q_{\mu}q^{\mu}}\geq0$.
Hence the NEC requires that 
\begin{equation}
\varepsilon_{u}+P\geq2q\,,\label{eq:NEC_q}
\end{equation}
which is a \emph{stronger} restriction than the one for a perfect fluid. Since the expressions $T_{\mu\nu}n^{\mu}n^{\nu}$ (with \emph{total} $T_{\mu\nu}$) and $\varepsilon_{u}+P-2q$ (for \emph{total} i.e.\ locally defined quantities) are four-scalars, they are invariant under coordinate transformations: their negativity in any one set of coordinates implies that the NEC is violated at this space-time point for all observers.

Since the universe is on average isotropic, the magnitude of the vector $q_{\mu}$ will generically be of the order of the perturbations. Moreover, in \emph{linear} perturbation theory  this vector $q_{\mu}$ is invariant under gauge and coordinate transformations because it does not have any background value, see e.g. \cite[Eq.~(7.14)]{Mukhanov:2005sc}. 

In general, a configuration of the perturbed fluid will feature non-zero perturbations $\delta\varepsilon$, $\delta P$ and $q_\mu$ on top of the background values, and where we have defined them in a comoving gauge with $\delta u^\mu=0$. In this case, the NEC is just
\begin{equation}
	\overline\varepsilon+ \overline{P} + \delta\varepsilon + \delta P \geq 2q\,,
\end{equation}
where overlines denote background quantities. Dependence on metric potentials is absent in this expression since it is purely a local measurement of a four-scalar quantity. 

On a background which just satisfies the NEC, with $w$ close to $-1$, these perturbations can be large enough to violate the condition \emph{locally}. For example, given an appropriate pressure perturbation, the local equation of state $w_\text{loc}=(\overline{P}+\delta P)/(\overline\varepsilon+\delta \varepsilon)$ can fluctuate sufficiently to do so. Violations of this type would be in principle thinkable even in a matter with an EMT of perfect-fluid form Eq.~\eqref{eq:EMTBackground}, although we stress every such model known so far is unacceptably unstable, see e.g.\ \cite{Dubovsky:2005xd}, and therefore unphysical.

Furthermore, \emph{only} in an imperfect fluid, there exist additional configurations in which $w_{\text{loc}}>-1$, and yet the NEC is violated,
\[
0<1+w_\text{loc}<\frac{2q}{\varepsilon}\,.
\]
A naive observer performing measurements of the cosmological background, or even of the pressure and energy density of the perturbations,
would from $w>-1$ conclude that the NEC is satisfied and the universe
is free from the negative energies discussed in the previous section.
However, when the perturbations are taken into account, it turns out
that the NEC is secretly violated. On the other hand, the knowledge
of $w$ and requirement of the NEC puts an upper bound on $q$ in
this situation.

We have expressed everything up to now in comoving gauge. One can appropriately derive the condition in other gauges and $T_{\mu\nu} n^\mu n^\nu$ will not necessarily have the same value. This is not worrying, but purely a result of the fact that gauge transformations change the space-time point assigned to particular coordinates. By changing a gauge, we are investigating the violation of the NEC in a neighboring space-time point. Working in any one fixed gauge is enough to ascertain whether the NEC is violated at some point in the universe.

\begin{figure}[t]
\begin{centering}
\includegraphics[width=0.92\columnwidth]{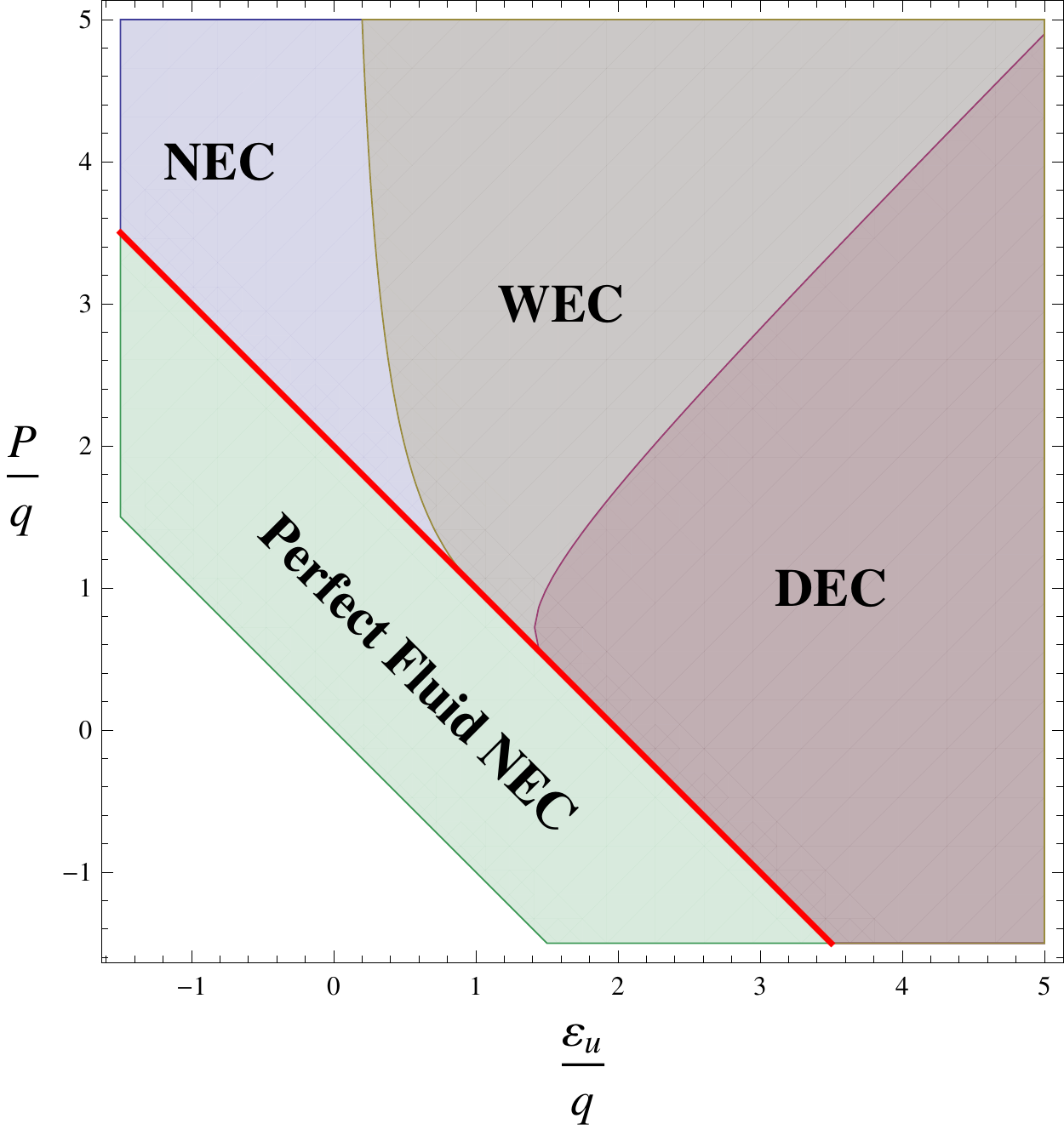}
\par\end{centering}

\caption{\label{fig:Energy} The shaded regions above the thick red line correspond
to DE configurations where various energy conditions are satisfied
in the space of energy density $\varepsilon_{u}$ and total pressure
$P$ normalised in units of energy flow $q=\sqrt{-q_{\mu}q^{\mu}}$.
The Dominant Energy Condition (DEC) is the strongest of the conditions
discussed in this paper and is satisfied in the region denoted by
DEC. The DEC implies the Weak Energy Condition (WEC), while the WEC
implies the Null Energy Condition (NEC). Hence the WEC holds in the
regions denoted DEC and WEC whereas NEC holds in the whole shaded
region above the red line. In the case of the perfect fluid the NEC
is weaker and holds in the whole shaded region including the shaded
part below the thick red line. }
\end{figure}

Interestingly the requirements from other energy conditions are also
stronger in the presence of the energy flow $q$, so that by observing
only $w$ one can easily be  misled. Below we provide the analytical
form of the conditions, which we have plotted for convenience in Figure
\ref{fig:Energy}; for discussion and derivation see \cite{Kolassis}.
To compare these conditions with the one for a canonical scalar field
or a perfect fluid see e.g.\ \cite{Carroll:2003st}.

The Weak Energy Condition (WEC), $T_{\mu\nu}V^{\mu}V^{\nu}\geq0$
for all time-like future-directed vectors $V^{\mu}$, is stronger
than NEC and is also important in various fundamental theorems in
GR. Below we formulate restrictions on $q$, $\varepsilon_{u}$ and
$P$ necessary for the WEC to hold. Provided the NEC holds there now
exists a minimal energy density that can be seen by any observer,
\begin{equation}
\varepsilon_{\text{min}}=\frac{\varepsilon_{u}-P}{2}+\sqrt{\left(\frac{\varepsilon_{u}+P}{2}\right)^{2}-q^{2}}\,.\label{eq:MinimalEnergyText}
\end{equation}
The WEC is equivalent to the condition $\varepsilon_{\text{min}}>0$
on top of the NEC (\ref{eq:NEC_q}). These inequalities can be also
written in the form of the following two options 
\begin{align*}
\text{either} &  & \varepsilon_{u}>q\,, &  & \text{and} &  & \varepsilon_{u}+P\geq2q\,,\\
\text{or} &  & q\geq\varepsilon_{u}>0\,, &  & \text{and} &  & \varepsilon_{u}P\geq q^{2}\,.
\end{align*}
 The observer measuring $\varepsilon_{\text{min}}$ is boosted against
the energy flow $q^{\mu}$ with the speed 
\[
v_{\text{min}}=\frac{\varepsilon_{u}+P}{2q}-\sqrt{\left(\frac{\varepsilon_{u}+P}{2q}\right)^{2}-1}\,.
\]

The frame of this observer,$V_{\mu}\propto u_{\mu}+v_{\text{min}}q_{\mu}/q$,
is the Landau-Lifshitz frame. If NEC is violated there is no Landau-Lifshitz
frame comoving with the energy (for a discussion of the choice of
cosmological frames see e.g. \cite{Sawicki:2012re}).The Dominant
Energy Condition (DEC): assumes that the WEC holds and that the density
of the fluid's 4-momentum $p_{\mu}=T_{\mu\nu}V^{\nu}$ is non-spatial:
$p_{\mu}p^{\mu}\geq0$ and future directed for all timelike and null
future-directed vectors $V^{\mu}$. At the end these conditions can
be written as the WEC plus on top of that $\varepsilon_{u}>q$ and
either $\varepsilon_{u}-P>q$ or 
\begin{align*}
 &  & q\geq\varepsilon_{u}-P\geq-q\,, &  & \text{and} &  & P\left(\varepsilon_{u}-P\right)\geq\frac{1}{2}q^{2}\,.
\end{align*}
 In particular, from these conditions it follows that the DEC always
implies that 
\[
\varepsilon_{u}\geq\sqrt{2}q\,,
\]
see Figure \ref{fig:Energy}. 
\begin{acknowledgments}
It is a pleasure to thank Asimina Arvanitaki, Fedor Bezrukov, Savas
Dimopoulos, Valerio Marra, Slava Mukhanov, Subodh Patil, Leonardo Senatore, Glenn Starkman and Wessel Valkenburg for very useful and valuable discussions. IS is supported by the DFG through
TRR33 ``The Dark Universe''. The work of AV is supported by ERC
grant BSMOXFORD no. 228169. IS would like to thank the CERN Theory
Group for hospitality during the preparation of this paper.
\end{acknowledgments}
\bibliographystyle{utcaps}
\bibliography{KGB}

\end{document}